\documentclass[12pt,nofootinbib,preprint,superscriptaddress]{revtex4}
\pdfoutput=1

\usepackage{amsmath}
\usepackage{amssymb}
\usepackage{anysize}
\usepackage{bold-extra}
\usepackage{color}
\usepackage{enumerate}
\usepackage{fancyhdr}
\usepackage{graphicx}
\usepackage[linktocpage,colorlinks,urlcolor=blue]{hyperref}
\usepackage{mathrsfs}
\usepackage{multirow}
\usepackage[final]{pdfpages}
\usepackage{rotating}
\usepackage{subfig}
\usepackage{textcomp}
\usepackage{url}
\usepackage{verbatim}
\usepackage{wrapfig}
\usepackage{amsfonts}
\usepackage{comment}
\usepackage{epigraph}
\usepackage[utf8]{inputenc}
\usepackage{slashed}
\usepackage{bbm}
\def\nn{\nonumber}

\def\gsim{\mbox{\raisebox{-.6ex}{~$\stackrel{>}{\sim}$~}}}
\newcommand{\be} {\begin{equation}}
\newcommand{\ee} {\end{equation}}
\newcommand{\ba} {\begin{eqnarray}}
\newcommand{\ea} {\end{eqnarray}}

\newcommand{\V}{\boldsymbol{V}}

\begin{document}

\title{B-decay Anomalies in a  Composite Leptoquark Model }

\author{Riccardo Barbieri}
\email{riccardo.barbieri@sns.it}
\affiliation{Scuola Normale Superiore, Piazza dei Cavalieri 7, 56126 Pisa, Italy and INFN, Pisa, Italy}

\author{Christopher W. Murphy}
\email{christopher.murphy@sns.it}
\affiliation{Scuola Normale Superiore, Piazza dei Cavalieri 7, 56126 Pisa, Italy and INFN, Pisa, Italy}
\affiliation{Department of Physics, Brookhaven National Laboratory, Upton, N.Y., 11973, U.S.A.}

\author{Fabrizio Senia}
\email{fabrizio.senia@sns.it}
\affiliation{Scuola Normale Superiore, Piazza dei Cavalieri 7, 56126 Pisa, Italy and INFN, Pisa, Italy}

\begin{abstract}
The collection of a few anomalies in semileptonic $B$-decays, especially in $b\rightarrow c \tau \bar{\nu}$, invites to speculate about the emergence of some striking new phenomena, perhaps interpretable in terms of a weakly broken $U(2)^n$ flavor symmetry and of leptoquark mediators. Here we aim at a partial UV completion of this interpretation by generalizing the minimal composite Higgs model to include a composite vector leptoquark as well.
\end{abstract}

\maketitle

\section{Introduction}
A number of anomalies in the decays of $B$ mesons continue to receive much attention.
As recalled below, the statistically most significant among these anomalies is of special interest since, at the partonic level, $b\rightarrow c \tau \bar{\nu}$, it involves three third generation particles. As such it is suggestive of an explanation in terms  of a $U(2)^n$ flavor symmetry that distinguishes between the third generation of fermions as singlets and the first two generations as doublets\cite{Barbieri:2012uh}.

Within this context ref.~\cite{Barbieri:2015yvd} looked at the ability of leptoquark models, in particular spin-one leptoquarks, to explain some of these anomalies.
However a model with massive vector fields cries out for a UV completion (see e.g.~\cite{Biggio:2016wyy}). This is particularly true since, as  one can anticipate from the relatively large size of the putative deviation from the Standard Model (SM) tree level amplitude, a fairly large coupling of the leptoquark must be invoked.
The aim of this paper is to investigate whether it is possible to make a composite model that can serve as a (more) UV complete explanation of these flavor anomalies. 
In particular, we are looking to generalize the simplified Composite Higgs Models (CHM) of~\cite{Contino:2006nn} to a case which includes leptoquarks. 
To this end we extend the global symmetry group of the strong sector from $SU(3) \times SO(5) \times U(1)~$\cite{Agashe:2004rs} to $SU(4) \times SO(5) \times U(1)$, where $SU(4)$ is the Pati-Salam group. The extension from $SU(3)$ to $SU(4)$ can be seen as natural if one thinks of composite leptons as necessary to give masses to the standard leptons by bilinear mixing, as in the quark case often discussed.

The experimental measurements of interest include a combined $4.0 \sigma$ excess over the SM, which is seen by three experiments in the charged current process
\begin{equation}
R_{D^{(*)}} = \frac{\Gamma(\bar{B} \to D^{(*)} \tau^- \bar{\nu}_{\tau})}{\Gamma(\bar{B} \to D^{(*)} \ell^- \bar{\nu}_{\ell})} ,
\end{equation}
with $\ell = e, \mu$.
Assuming a common scaling of $R_D$ and $R_{D^*}$ with respect to their SM predictions, a one parameter fit to the averages presented by the Heavy Flavor Averaging Group (HFAG) yields $R_{D^{(*)}} / (R_{D^{(*)}})_{\text{SM}} = 1.27 \pm 0.06$~\cite{Amhis:2014hma, HFAG}. 
The HFAG result makes use of experimental measurements from BaBar~\cite{Lees:2012xj, Lees:2013uzd}, LHCb~\cite{Aaij:2015yra}, and Belle~\cite{Huschle:2015rga, Abdesselam:2016cgx} (see also~\cite{Sato:2016svk}); as well as the theoretical predictions of refs.~\cite{Fajfer:2012vx, Na:2015kha} (see also~\cite{Kamenik:2008tj, Lattice:2015rga}).

Furthermore, LHCb has reported~\cite{Aaij:2013qta, Aaij:2014ora} a $2.6\sigma$ deviation from the SM in the neutral current process
\begin{equation}
R_K = \frac{\Gamma(B^+ \to K^+ \mu^+ \mu^-)}{\Gamma(B^+ \to K^+ e^+ e^-)} ,
\end{equation}
possibly indicating a violation of lepton flavor universality (LFU).
Specifically, for $M_{\ell^+ \ell^-}^2 \in [1,\, 6]~\text{GeV}^2$ the measured value of $R_K$ is $0.745_{-0.074}^{+0.090} (\text{stat}) \pm 0.035 (\text{syst})$, compared to a SM value that is close to 1~\cite{Bobeth:2007dw, Bordone:2016gaq}. Global fits to all $b \to s \ell \ell$ data seem to indicate a more general tension with the SM~\cite{Altmannshofer:2014rta, Descotes-Genon:2015uva}.\footnote{A recent update of~\cite{Altmannshofer:2014rta} claims the combined tension with the SM has increased to $4.5\sigma$~\cite{Straub}.} However, many of these observables are subject to significant hadronic uncertainties, whereas $R_K$ and $R_{D^{(*)}}$, are not. 

On the theoretical side of things, both dynamical models~\cite{Datta:2013kja, Hiller:2014yaa, Gripaios:2014tna, Sahoo:2015wya, Becirevic:2015asa, Calibbi:2015kma, Freytsis:2015qca, Bauer:2015knc, Fajfer:2015ycq, Becirevic:2016oho} and bottom-up effective field theory approaches~\cite{Altmannshofer:2014rta, Bhattacharya:2014wla, Alonso:2015sja, Descotes-Genon:2015uva} have been, and continue to be used to analyze and explain the experimental results.
The dynamical models typically involve some kind of leptoquarks, but not always, see ref.~\cite{Greljo:2015mma}. 
More recent work on this topic often focuses on other potential experimental signatures of these anomalies and models, including but not limited to: kaon physics~\cite{Crivellin:2016vjc, Kumar:2016omp, Coluccio-Leskow:2016tsp}, kinematic distributions in the decays of $B$ mesons~\cite{Becirevic:2016zri, Alonso:2016gym, Ligeti:2016npd}, tau lepton searches~\cite{Buttazzo:2016kid, Faroughy:2016osc}, dark matter~\cite{Sierra:2015fma, Altmannshofer:2016jzy}, and the evolution of the renormalization group equations (RGE) that leads to multiple effects~\cite{Feruglio:2016gvd}.
 
Ref.~\cite{Feruglio:2016gvd} is especially of interest in light of the goal of this work as it challenges the idea that the $B$ decay anomalies could be due to simple extensions of the SM, i.e. a single leptoquark field. 
Specifically, when only the minimal set of operators needed to explain $R_{D^{(*)}}$ and $R_K$ are generated at some scale $\Lambda \gg v$, the RG evolution of these operators generates unacceptably large deviations from lepton flavor universality in $Z$ and $\tau$ decays as well as lepton flavor violating $\tau$ decays.
The particular operators are $Q_{\ell q}^{(1)}$ and $Q_{\ell q}^{(3)}$; see~\cite{Jenkins:2013wua, Alonso:2013hga} for notation and the explicit form of the RGE.
While a full one-loop RGE analysis is beyond the scope of this work, we note there are at least two effects that distinguish the model under consideration in this work from that of ref.~\cite{Feruglio:2016gvd}.
The first is that there are more dimension-six operators than the two listed above, which are generated at tree level that contribute to the relevant RGE, e.g. $Q_{\ell \ell}$ and $Q_{H \Box}$, as well operators that do not contribute to the RGE of interest.
Some of the additional operators contribute to the RGE with the opposite sign of the contribution coming from the operators considered in~\cite{Feruglio:2016gvd}.
Secondly there are direct contributions to the observables of interest that are generated at the scale $\Lambda$ at the one-loop level.
Though these contributions do not have a log-enhancement as the RGE contributions do, they can still serve to partially cancel the effects of the RGE contributions.

The rest of this paper is organized as follows. 
Sec.~\ref{sec:PC} describes the field content of the model as well as the mass spectra and mixing angles associated with the fermions and vector bosons. 
The tree level amplitudes and viable parameter space are presented in sec.~\ref{tran}. 
This is followed by a discussion of electroweak precision data in sec.~\ref{sec:ewpt}. 
Then in sec.~\ref{sec:dist} a description is given of a number of features of this model, which distinguish it from the usual partial CHM. 
Finally, our conclusions are given in sec.~\ref{sec:conc}.

\section{Particle Content}
\label{sec:PC}
We start by describing the field content of the composite sector in terms of its representations under the unbroken global symmetry of the composite sector, $SU(4) \times SU(2)_L \times SU(2)_R \times U(1)_X$. 
The composite Higgs, $\mathcal{H} = (H,\, \tilde{H})$ where $\tilde{H} = i \sigma^2 H^{\star}$, is a bidoublet of $SU(2)_{L,R}$. 
The hypercharge is given by
\begin{equation}
\label{eq:hype}
Y = \sqrt{\tfrac{2}{3}} T^{15} + T_R^3 + X,
\end{equation}
where $T^{A = 1, \ldots, 15}$ are the generators of $SU(4)$ with normalization $\text{Tr}(T^A T^B) = \delta^{AB} / 2$. 
The coefficient in front of $T^{15}$ in eq.~\eqref{eq:hype} is necessary to get the correct hypercharge, since $\sqrt{2/3}\, T^{15} = (B-L) / 2$, with $B$ and $L$ being baryon and lepton numbers respectively.

\subsection{Vector Boson Masses and Mixings}
\label{cvm}
The vector boson masses and mixings are analogous to those of ref.~\cite{Contino:2006nn} or to those of a two site model of the standard CHM, apart from two main differences: we have to include $SU(4)$ instead of $SU(3)$ and the elementary weak hypercharge gauge boson mixes  with three composite fields (associated with $T^{15}$, $T_{R}^{3}$ and $X$). 
The $SU(4)$ composite bosons can be written as
\begin{equation}
\label{4matrix}
\rho_{\mu} = \rho^A_{\mu} T^A = 
\begin{pmatrix}
\tfrac{1}{2} \rho_{\mu}^a \lambda^a + \tfrac{1}{2\sqrt{6}} \rho^{15}_{\mu} \mathbbm{1}_{3\times3} & \tfrac{1}{\sqrt{2}} V_{\mu} \\
\tfrac{1}{\sqrt{2}} V_{\mu}^{\dagger} & - \tfrac{3}{2\sqrt{6}} \rho^{15}_{\mu}
\end{pmatrix} ,
\end{equation}
where $\lambda^{a = 1, \ldots, 8}$ are the generators of $SU(3)$, and the leptoquarks, $V$ and $V^{\dagger}$, are associated with the $A = 9, \ldots, 14$ generators of $SU(4)$. 
The composite bosons in the adjoint of $SU(2)_L\times SU(2)_R \times U(1)_X$ are 
\begin{equation}
 \label{221composites}
 W_\mu^L=W_\mu^{L\alpha}T^{\alpha}_{L},\quad\quad  W_\mu^R=W_\mu^{R\alpha}T^{\alpha}_{R},\quad\quad V_{\mu}^X,
 \end{equation}
  respectively.
 $T_{L,R}^{\alpha=1,2,3}$ are the generators of $SU(2)_{L,R}$, with the same normalization of the $SU(4)$ generators. 
In general for the four group factors $\{SU(4), SU(2)_L, SU(2)_R, U(1)_X\}$ there are four  strong couplings $\{g_{\rho}, g_{\rho L},g_{\rho R},g_{X}\}$ and four masses $\{M_{\rho}, M_{\rho L},M_{\rho R},M_{X}\}$. 
With the only purpose of simplifying the formulae in the following we take $g_{\rho R} = g_X$ and $M_{\rho R} = M_X$.
On the other hand, before mixing, the elementary fields associated with the $SU(3)\times SU(2)\times U(1)$ gauge group are
\begin{equation}
\label{321elementary}
G_{e,\mu} = G_{e,\mu}^aT_3^a,\quad\quad 
W_{e,\mu} = W_{e,\mu}^\alpha T_L^\alpha, \quad\quad
B_{e,\mu},
\end{equation}
with their own couplings  $\{g_{e3}, g_{e2},g_{e1}\}$.

After mixing (and prior to electroweak symmetry breaking), the mass eigenstates are superpositions of the states in~\eqref{4matrix},~\eqref{221composites},~\eqref{321elementary}.
Of special interest to us are the leptoquarks, $V_\mu, V_\mu^{\dagger}$, which stay unmixed, and the totally neutral states, in number four,  one of which, $\tilde{V}_\mu = (W_\mu^{R3} - V^X_{\mu}) / \sqrt{2}$, stays also unmixed.\footnote{This is a feature of the simplifying choice $g_{\rho R}= g_X$.}
In the following we shall take
\begin{equation}
\frac{M_\rho^2}{g_\rho^2}=\frac{M_{\rho L}^2}{g_{\rho L}^2}=\frac{M_{\rho R}^2}{g_{\rho R}^2}= \frac{f^2}{2},
\end{equation}
and present results at leading order in  $\xi \equiv v^2 / f^2$, where $v= \sqrt{2}g_2 M_W\approx 175$ GeV.

Defining $B_\mu^* = (W_\mu^{R3} + V^X_{\mu}) / \sqrt{2}$ and calling $\rho_{c\mu}$ and $\rho_{e\mu}$ the collections of composite and elementary vectors respectively, the Lagrangian for the gauge sector is
\begin{align}
\label{eq:Lga}
\mathcal{L}_{\text{gauge}} 
&= 
- \frac{1}{4} \sum_{c}\rho_{c\mu\nu} \rho_{c}^{\mu\nu} -\frac{1}{4}\sum_{e}\rho_{e\mu\nu} \rho_{e}^{\mu\nu} \\
&+ 
 M_{\rho}^2V_\mu V^{\mu +} +  M_{\rho R}^2 W_{\mu R} W^{\mu R+}  +  \frac{1}{2}M_{\rho R}^2\tilde{V}_\mu^2 \nn \\
&+ 
 \frac{M_{\rho}^2}{2g_\rho^2}(g_{e3}  G_{e, \mu}^a - g_\rho \rho^a_\mu)^2 +  \frac{M_{\rho}^2}{2g_\rho^2}(\sqrt{2/3}g_{e1}  B_{e, \mu} - g_\rho \rho^{15}_ \mu)^2 \nn \\
 &+
 \frac{M_{\rho L}^2}{2g_{\rho L}^2}(g_{e2}  W_{e, \mu}^a - g_{\rho L} W^L_\mu)^2 +  \frac{M_{\rho R}^2}{2g_{\rho R}^2}(\sqrt{2}g_{e1}  B_{e, \mu} - g_{\rho R}B^*_ \mu)^2 . \nn
\end{align}
The mass eigenstates (prior to electroweak symmetry breaking) are related to the states in eq.~\eqref{eq:Lga} by
\begin{align}
\begin{pmatrix}
G_{e, \mu}^a \\
\rho_{\mu}^a
\end{pmatrix}
&= 
\begin{pmatrix}
\cos \theta_3 & - \sin \theta_3 \\
\sin \theta_3 &  \cos \theta_3
\end{pmatrix}
\begin{pmatrix}
G_{\mu}^a \\
G_{\mu}^{Ha}
\end{pmatrix}, \\
\begin{pmatrix}
W_{e, \mu}^{\alpha} \\
W_{\mu}^{L \alpha}
\end{pmatrix}
&= 
\begin{pmatrix}
\cos \theta_2 & - \sin \theta_2 \\
\sin \theta_2 &  \cos \theta_2
\end{pmatrix}
\begin{pmatrix}
W_{\mu}^{\alpha} \\
W_{\mu}^{H \alpha}
\end{pmatrix}, \nn \\
\begin{pmatrix}
B_{e, \mu} \\
B^*_{\mu}\\
\rho_{\mu}^{15}
\end{pmatrix}
&\approx \begin{pmatrix}
1 & - \sqrt{2}\frac{g_{e1}}{g_{\rho R}}& - \sqrt{\frac{2}{3}}\frac{g_{e1}}{g_{\rho }} \\
\sqrt{2}\frac{g_{e1}}{g_{\rho R}}&  1&\mathcal{O}(\frac{g_{e1}}{g_{\rho,\rho R}})^2 \\
\sqrt{\frac{2}{3}}\frac{g_{e1}}{g_{\rho }}&\mathcal{O}(\frac{g_{e1}}{g_{\rho,\rho R}})^2 &1
\end{pmatrix}
\begin{pmatrix}
B_{\mu} \\
B^{H}_{\mu}\\
X_{\mu}
\end{pmatrix}, \nn
\end{align}
where the mixing angles $\theta_2$ and $\theta_3$ are
\begin{equation}
\{\theta_2,\theta_3\}=\Big\{
\text{arctan}\Big(\frac{g_{e2}}{g_{\rho L}}\Big),\text{arctan}\Big(\frac{g_{e3}}{g_{\rho}}\Big)\Big\} .
\end{equation}
For simplicity we show the rotation matrix in the neutral sector at leading order in
$  \frac{g_{e1}}{g_{\rho R}}\ll1$ and 
$\frac{g_{e1}}{g_{\rho }}\ll 1 $. 
The physical gauge couplings are $g_3 = g_{e 3} c_3 = g_{\rho} s_3$, $g_2 = g_{e 2} c_2 = g_{\rho L} s_2$ and 
$g_1 \approx g_{e1}$.

The mass spectrum in the gauge sector is
\begin{align}
\mathcal{L}_{\text{gauge}} &\supset \frac{1}{2} \frac{M_{\rho}^2}{c_3^2} G_{\mu}^{Ha} G^{ H a \mu} + M_{\rho}^2 V_{\mu}^{\dagger} V^{\mu} + \frac{1}{2} \frac{M_{\rho}^2}{c_{15}^2} X_{\mu} X^{\mu} \\
&+
\frac{1}{2} \frac{M_{\rho L}^2}{c_2^2} W_{\mu}^{Ha} W^{ H a \mu} + M_{\rho_{R}}^2 W_{\mu}^{R +} W^{R - \mu} + \frac{1}{2} M_{\rho R}^2 
\tilde{V}_\mu \tilde{V}^\mu + \frac{1}{2} \frac{M_{\rho R}^2}{c_R} B^H_{\mu} B^{H \mu} , \nn
\end{align}
where $c_{2,3} = \cos \theta_{2,3}$, $c_{15} \approx 1+\mathcal{O}(\frac{g_{e1}}{g_{\rho }} )^2$ and 
$c_R \approx 1+\mathcal{O}(\frac{g_{e1}}{g_{\rho R }} )^2$. Since also $\theta_{2,3}$ are bound to be small, one can consider to a reasonable approximation only three different masses involved, or in fact only two after imposing $g_{\rho R}=g_{\rho L}$ to respect custodial symmetry
\begin{equation}
M_{G^H}\approx M_V\approx M_X \approx M_{\rho}
\end{equation}
and
\begin{equation}
M_{W^R}\approx M_{W^H}\approx M_{\tilde{V}} \approx M_{B^H}\approx M_{\rho R}.
\end{equation}
Note that
\begin{equation}
\label{eq:vecmass}
M_{W^H} = \frac{M_W}{\sqrt{\xi} s_2 c_2}.
\end{equation}
The part of the Lagrangian that describes the leptoquark interactions with the SM gauge fields is given by
\begin{align}
\label{eq:LagV}
\mathcal{L}_V &= \left(D_{\mu} V_{\nu}\right)^{\dagger} \left(D^{\nu} V^{\mu}\right) - \left(D_{\mu} V_{\mu}\right)^{\dagger} \left(D^{\mu} V^{\nu}\right) + M_{\rho}^2 V_{\mu}^{\dagger} V^{\mu} \\
&- i g_3 G^{a \mu\nu} \left(V_{\mu}^{\dagger} \frac{\lambda^a}{2} V_{\nu}\right) - i g_1 Y B^{\mu\nu} \left(V_{\mu}^{\dagger} V_{\nu}\right) , \nn
\end{align}
where $D_{\mu} = \partial_{\mu} - i g_3 \frac{\lambda^a}{2} G_{\mu}^a - i g_1 Y B_{\mu}$ with $Y = 2/3$.
From the above interaction terms one finds that the leptoquark $V_{\mu}$ couples to the
SM fields $G_{\mu}^a$ and $B_\mu$ as in \cite{Barbieri:2015yvd} with $k_s = k_Y = 1$.
This means that potentially dangerous contributions to dipole operators (responsible for $\mu \rightarrow e \gamma$, $\tau \rightarrow \mu \gamma$ decays, etc.) are finite in our model, unlike in the general case of the low energy leptoquark model \cite{Barbieri:2015yvd}. 

\subsection{Fermion Masses and Mixings}
We want to extend the so-called bidoublet  model\footnote{We adopt the nomenclature of ref. \cite{Barbieri:2012tu}.}   commonly considered in the standard Composite Higgs picture to the case of $SU(4)$. 
The triplet scenario can be dealt with in a similar way and is discussed in appendix~\ref{sec:trip}.
The composite fermions transform under  $SU(4)\times SU(2)_L \times SU(2)_R \times U(1)_X$ as $\psi_{\pm}=(4,2,2)_{\pm 1/2}$ and $\chi_{\pm}=(4,1,1)_{\pm 1/2}$.

Our notation for the bidoublet model is
\begin{equation}
\psi_+ = 
\begin{pmatrix}
Q_+^{\beta} \\
L_+
\end{pmatrix}, \quad
\psi_- = 
\begin{pmatrix}
Q_-^{\beta} \\
L_-
\end{pmatrix}, \quad
\chi_+ = 
\begin{pmatrix}
\tilde{U}^{\beta} \\
\tilde{N}
\end{pmatrix}, \quad
\chi_- = 
\begin{pmatrix}
\tilde{D}^{\beta} \\
\tilde{E}
\end{pmatrix},
\end{equation}
where $\beta = 1,2,3$ is a fundamental color index. 
The components of $\psi_{\pm}$ are further reduced as
\begin{align}
Q_+ &=
\begin{pmatrix}
Q^{(U)} & X^{(U)}
\end{pmatrix}
=
\begin{pmatrix}
U & X^{5/3} \\
D & X^{2/3}
\end{pmatrix}, \\
L_+ &=
\begin{pmatrix}
L^{(N)} & X^{(N)}
\end{pmatrix}
=
\begin{pmatrix}
N & X^{+1} \\
E & X^{0}
\end{pmatrix}, \nn \\
Q_- &=
\begin{pmatrix}
X^{(D)} & Q^{(D)}
\end{pmatrix}
=
\begin{pmatrix}
X^{-1/3} & U^{\prime} \\
X^{-4/3} & D^{\prime}
\end{pmatrix}, \nn \\
L_- &=
\begin{pmatrix}
X^{(E)} & L^{(E)}
\end{pmatrix}
=
\begin{pmatrix}
X^{-1} & N^{\prime} \\
X^{-2} & E^{\prime}
\end{pmatrix}. \nn
\end{align}
where in the right-hand side of these equations we make explicit the transformation properties of the various components under the SM gauge group. All the $X$ states are exotic with their charge explicitly indicated, while their $SU(3)$ properties are left understood.

Following ref.~\cite{Barbieri:2012tu}  we attribute the basic distinction between the third and the lighter first and second generations to the presence of an approximate $U(2)^n$ flavor symmetry which is unbroken in the composite sector and is weakly broken along specific ``spurion'' directions only in the mass mixings between the elementary and the composite fermions. In particular, to avoid unobserved flavor-breaking effects, we rely on the idea of left- or right-compositeness~\cite{Cacciapaglia:2007fw, Barbieri:2008zt,Redi:2011zi}.
In the present context left-compositeness and right-compositeness can be implemented invoking as intermediate symmetries
\begin{equation}
\mathcal{G}_{LC}=U(2)_{q+l+\psi_{\pm}+\chi_{\pm}}\times U(2)_u \times
U(2)_d\times U(2)_e ,
\end{equation}
or
\begin{equation}\mathcal{G}_{RC}=
U(2)_q \times U(2)_l\times U(2)_{u+\psi_++\chi_+} \times U(2)_{d+e+\psi_-+\chi_-} ,
\end{equation}
respectively. 
In particular, in the case of right-compositeness, one ends up with flavor violation in the up quark sector suppressed by inverse power of $z_3\equiv s_{Lu3} / s_{Ld3}$, as defined below, which is required to be large by consistency with the $Zb_L \bar{b}_L$ 
coupling measurements (see sec.~\ref{sec:ewpt}). 
This, in turn, suppresses the contribution to the  charged-current $B$ anomaly, making impossible to reproduce the observed deviation. 
Therefore, in the following we will consider only  left-compositeness.

The Yukawa and mass terms for the fermionic resonances in the strong sector are given by
\begin{equation}
\begin{split}
\mathcal{L}^{\text{bidoublet}}_{s}=
m_{\psi_+}\text{Tr}[\bar{\psi}_+ \psi_+]+m_{\psi_-}\text{Tr}[\bar{\psi}_- \psi_-]+
m_{\chi_+}(\bar{\chi}_+ \chi_+)+m_{\chi_-}(\bar{\chi}_-\chi_-)+\\
\Big(
Y_+^{ii}\, \text{Tr}[\bar{\psi}_+^i \mathcal{H}]_L \chi_{R +}^i + 
Y_-^{ii}\, \text{Tr}[\bar{\psi}_-^i \mathcal{H}]_L \chi_{R -}^i + \text{h.c.} \Big) ,
\end{split}
\end{equation}
where $Y_{\pm}$, $m_{\psi_{\pm}}$ and $m_{\chi_{\pm}}$ are $U(2)$ preserving flavor diagonal matrices, so that $Y^T_\pm = (Y_{\pm 3} , Y_{\pm 2} , Y_{\pm 2})$, and similarly for $m_{\Psi \pm}$ and $m_{\chi \pm}$.
As in ref.~\cite{Barbieri:2012tu} the quark mixing Lagrangian is given by 
\begin{align}
\label{mixing2L}
 \mathcal{L}_\text{q mix,LC}^{\text{bidoublet}} &=
m_{\psi_{+3}}\lambda_{Lu3}\bar{q}_{L3}U_{3R} +
m_{\psi_{+2}}\lambda_{Lu2} \mathbf{\bar{q}_L} \mathbf{U_R} +
m_{\chi_{+3}}\lambda_{Ru3}\tilde{U}_{3L} t_R \\
&+m_{\chi_{+2}}\,d_u\, (\bar{\tilde{\mathbf{U}}}_\mathbf{L}\V)t_R  +
m_{\chi_{+2}}\, \bar{\tilde{\mathbf{U}}}_\mathbf{L} \Delta_u \bar{\mathbf{u}}_{\mathbf{R}} +
\text{h.c.}
+ (u,\tilde{U},t,\tilde{U}_3,+\rightarrow d,\tilde{D},b,\tilde{D}_3,-) . \nonumber
\end{align}
Similarly the lepton mixing Lagrangian is
\begin{align}
\label{mixinge}
  \mathcal{L}_\text{e mix,LC}^{\text{bidoublet}} &=
m_{\psi_{-3}}\lambda_{Le3}\bar{l}_{3L}L^{(E)}_{3R} +
m_{\psi_{-2}}\lambda_{Le2} \bar{\mathbf{l}}_{\mathbf{L}}\mathbf{L^{(E)}_R} +
m_{\chi_{-3}}\lambda_{Re3}\bar{\tilde{E}}_{3L}\tau_R \\
&+m_{\chi_{-2}}\,d_e\, (\bar{\tilde{\mathbf{E}}}_\mathbf{L}\V)\tau_R +
m_{\chi_{-2}}\bar{\tilde{\mathbf{E}}}_\mathbf{L} \Delta_e \bar{\mathbf{e}}_{\mathbf{R}} +
\text{h.c.} . \nonumber
\end{align}
The mixings in the first lines of~\eqref{mixing2L} and~\eqref{mixinge} break the symmetry of the strong sector down to $\mathcal{G}_{LC}$. 
This symmetry is in turn broken minimally by the spurions
\begin{align}
\V \sim (2,1,1,1), \quad\quad \Delta_u \sim (2,2,1,1), \quad\quad \Delta_d \sim (2,1,2,1), \quad\quad \Delta_e \sim (2,1,1,2) ,
\label{MU2spurionssu4}
\end{align}
in the second lines of the same equations.

The SM Yukawa couplings for up and down quarks can be written in terms of the spurions as in \cite{Barbieri:2012tu}. Adopting also the same definitions as in \cite{Barbieri:2012tu} for the mixings $s_L, s_R$ between the elementary and the composite fermions, it is
\begin{align}
\label{SMYukLC}
\hat y_u &= \begin{pmatrix}a_u\, \Delta_u & y_t\V\\ 0 & y_t\end{pmatrix}, &
\hat y_d &= \begin{pmatrix}a_d\, \Delta_d &y_b x_b\V\\ 0 & y_b\end{pmatrix},
\end{align}
while for the charged lepton we obtain
\begin{align}
\label{SMYukLCe}
\hat y_e &= 
\begin{pmatrix}
a_e\, \Delta_e & y_\tau x_\tau \V\\ 0 & y_\tau 
\end{pmatrix} ,
\end{align}
where
\begin{align}
y_t &= Y_{+3}s_{Lu3}s_{Ru3},
\\
a_u &= Y_{+2}s_{Lu2},& x_b &=\Big(\frac{d_d Y_{-2}Y_{+3}}{d_u  Y_{-3}Y_{+2}}\Big)
\frac{s_{Ld2}s_{Lu3}t_{Ru3}}{s_{Lu2}s_{Ld3}t_{Rd3}},
& &  \\
y_\tau &= Y_{-3}s_{Le3}s_{Re3},
\\
a_e &= Y_{-2}s_{Le2},& x_\tau &=\Big(\frac{d_e Y_{-2}Y_{+3}}{d_u  Y_{-3}Y_{+2}}\Big)
\frac{s_{Le2}s_{Lu3}t_{Ru3}}{s_{Lu2}s_{Le3}t_{Re3}},
\end{align}
and similarly for $y_b$ and $a_d$ with the obvious replacements.
Extending $\mathcal{G}_{LC}$ by $U(1)_{e3}\times U(1)_{d3}$ can explain the smallness of $y_\tau$ and $y_b$, making natural room for $s_{R(e,d)3}\ll s_{Ru3}$ and $d_{e,d} \ll d_{u}$. 

The Yukawa matrices $\hat{y}_{u,d,e}$ are diagonalized to a sufficient level of approximation by unitary transformations only acting on the left-handed quarks and leptons, $U_{u,d,e}$. This is why there are chirality-conserving flavor-changing interactions only among left-handed fermions.
The CKM matrix is given by  $V = U_u^{\dagger}U_d$. 
The parameter $x_b$ is not determined by CKM data and it enters in flavor violating terms in up and down quark sector via 
\begin{equation}
r_u=1/(1-x_b),\quad\quad r_d=1-r_u
\end{equation}
respectively. 

Here we are not concerned with neutrino masses and mixings, which can arise from a suitable Majorana mass matrix of the right handed neutrinos mixed with the composite $\tilde{N}$ states. 
In any event, to an excellent level of approximation, we can study $B$ anomalies in the basis of neutrino current-eigenstates, where the charged current leptonic weak interactions are flavor-diagonal. 

\section{Tree level amplitudes for B anomalies}
\label{tran}
Exchanges of spin-one resonances contribute to tree level $b\rightarrow c \tau \nu$ and $b\rightarrow s \ell \ell$ decays as well as to $\Delta F=2$ transitions. The interaction Lagrangian of the composite vectors with the elementary quarks and leptons in the mass basis is given in appendix~\ref{sec:vecferm}.
We shall neglect terms suppressed by $1/z_3=s_{Ld3}/s_{Lu3}$ and $1/z_{3e}=s_{Le3}/s_{L\nu 3} $ as $z_3, z_{3e}$ are required to be large to control the deviations from the SM of the 
$Z b_L \bar{b}_L $ and $Z \tau_L \bar{\tau}_L $ couplings respectively. 
It is also convenient to define the following quantity
\begin{equation}
\label{eq:Adef}
 A^{(V)}_{f} \equiv \Big(1-\frac{t_V^2}{t^2_{f}}\Big), \text{where }
 V=\{2,3,15,R\}, \text{and } f=\{Lu3,L\nu3\} .
\end{equation}

Contributions to the operator $(\bar{c}_L \gamma_\mu b_L)(\bar{\tau}_L\gamma^\mu \nu_{3L})$ arise from the $t$-channel exchange of the leptoquark $V_{\mu}$ and the $s$-channel exchanges of $W^{H\pm}$.
For $b\rightarrow c  \tau  \bar\nu_3$ one has
\begin{equation}
\label{Leff_cc}
\mathcal{L}_{eff}^{b\rightarrow c\tau\nu}= 
r_u V_{cb}
\Big(-\frac{g_2^2}{M_W^2}\Big)[\xi s^2_{Lu3} s^2_{L\nu 3}] \Big(\frac{1}{2}+\frac{1}{2}f_{W*}\Big)
(\bar{c}_L\gamma_\mu b_L)(\bar{\tau}_L\gamma_\mu \nu_{3L}) ,
\end{equation}
where
\begin{equation}
\label{eq:f}
 f_{W*}(\theta_2,\theta_{Lu3}, \theta_{L\nu 3})\equiv 
 c_2^4 A^{(2)}_{Lu3} A^{(2)}_{L\nu3} .
\end{equation}
For small $\theta_2$, $f_{W*}$ tends towards one, so that
 to explain the $R_{D^{(*)}}$ anomaly at $1\sigma$ one needs
\begin{equation}
\label{eq:RD}
s_{L u 3}^2 s_{L \nu 3}^2 = (0.49 \div 0.77) \left(\frac{1.00}{r_u}\right) \left(\frac{0.10}{\xi}\right).
\end{equation}

For the neutral current process $b\rightarrow s\mu\mu$, there are leading contributions from the $t$-channel exchange of the leptoquark $V_\mu$ and the $s$-channel exchanges of $W^{H3}, \tilde{V}$ and $X$.
 One finds
\begin{align}
\label{Leff_nn}
\mathcal{L}_{eff}^{b\rightarrow s\mu\mu} &= 
r_d V_{ts}(c_l \epsilon_l)^2
\Big(-\frac{g_2^2}{M_W^2}\Big)[\xi s^2_{Lu3} s^2_{L\nu 3}] \\
&\times \Big(1+
\frac{1}{4}f_{W*}-\frac{1}{12}\frac{3}{2}f_X\Big)
(\bar{s}_L\gamma_\mu b_L)(\bar{\mu}_L\gamma_\mu \mu_{L}) , \nn
\end{align}
where $f_X=f_{W^*}$ with $\theta_2 \rightarrow \theta_{15}$.
Note that the other neutral vector $B^H$ has couplings to the SM fermions that vanish as $(g_{e1}/g_{\rho R})^2$ from to the elementary fermionic current of $B_e$, whereas $B^*$ couples only with exotic states ($X^{2/3}$, $X^{5/3}$, $X^{-1/3}$, $X^{-4/3}$,$\dots$).
At zeroth order in the mixing angles of the gauge sector, in order to reproduce the neutral current anomaly one needs, based on the best-fit values from~\cite{Altmannshofer:2014rta}, 
\begin{equation}
 s^2_{Lu3} s^2_{L\nu 3}
\approx
- (0.65 \div 1.31) \Big(\frac{0.07}{c^2_l\epsilon^2_l}\Big)\Big(\frac{0.04}{r_d}\Big)\Big(\frac{0.10}{\xi}\Big) .
\end{equation}
\begin{figure}
   \centering
\subfloat{\includegraphics[width=0.48\textwidth]{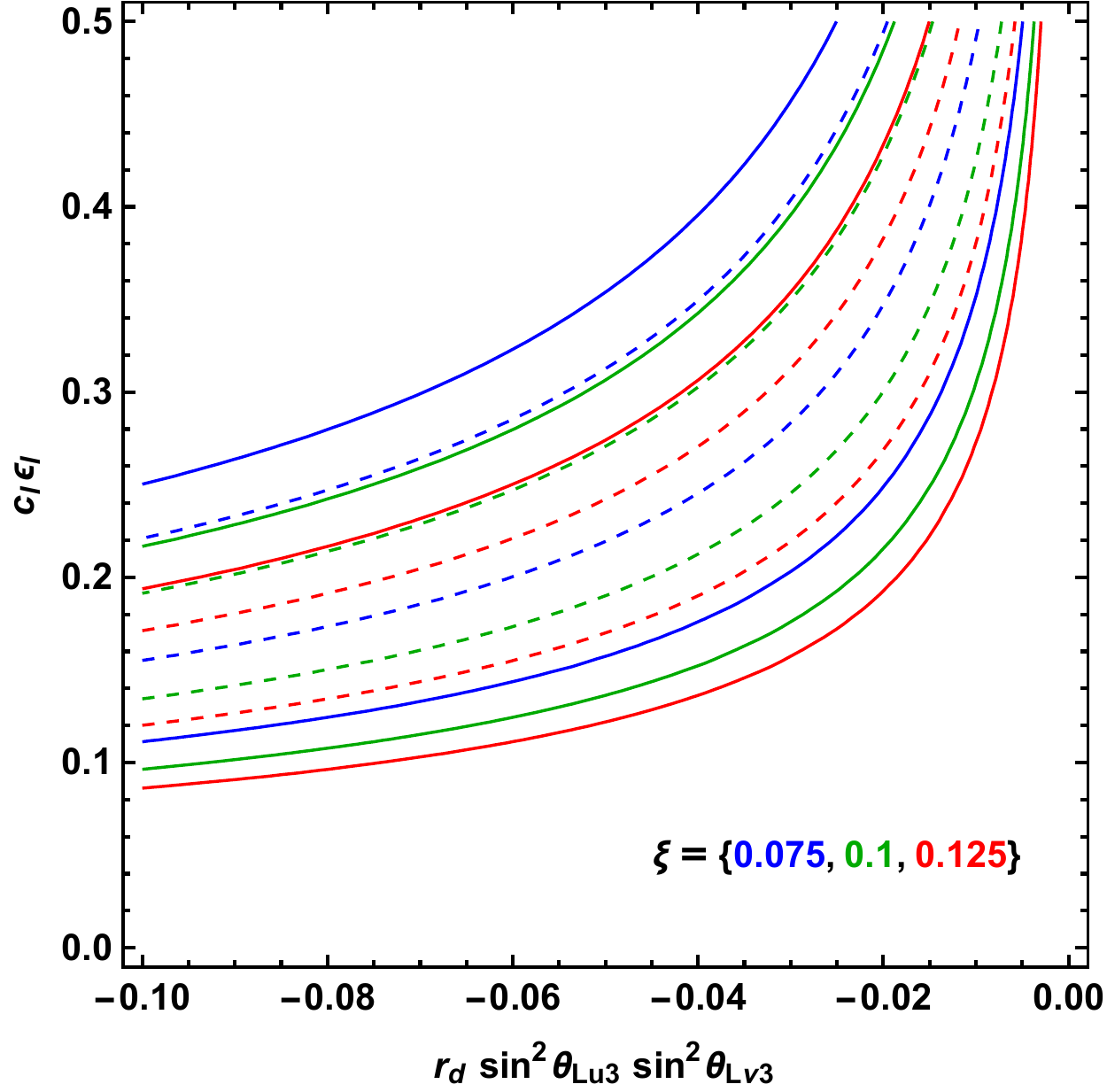}}\,
\subfloat{\includegraphics[width=0.48\textwidth]{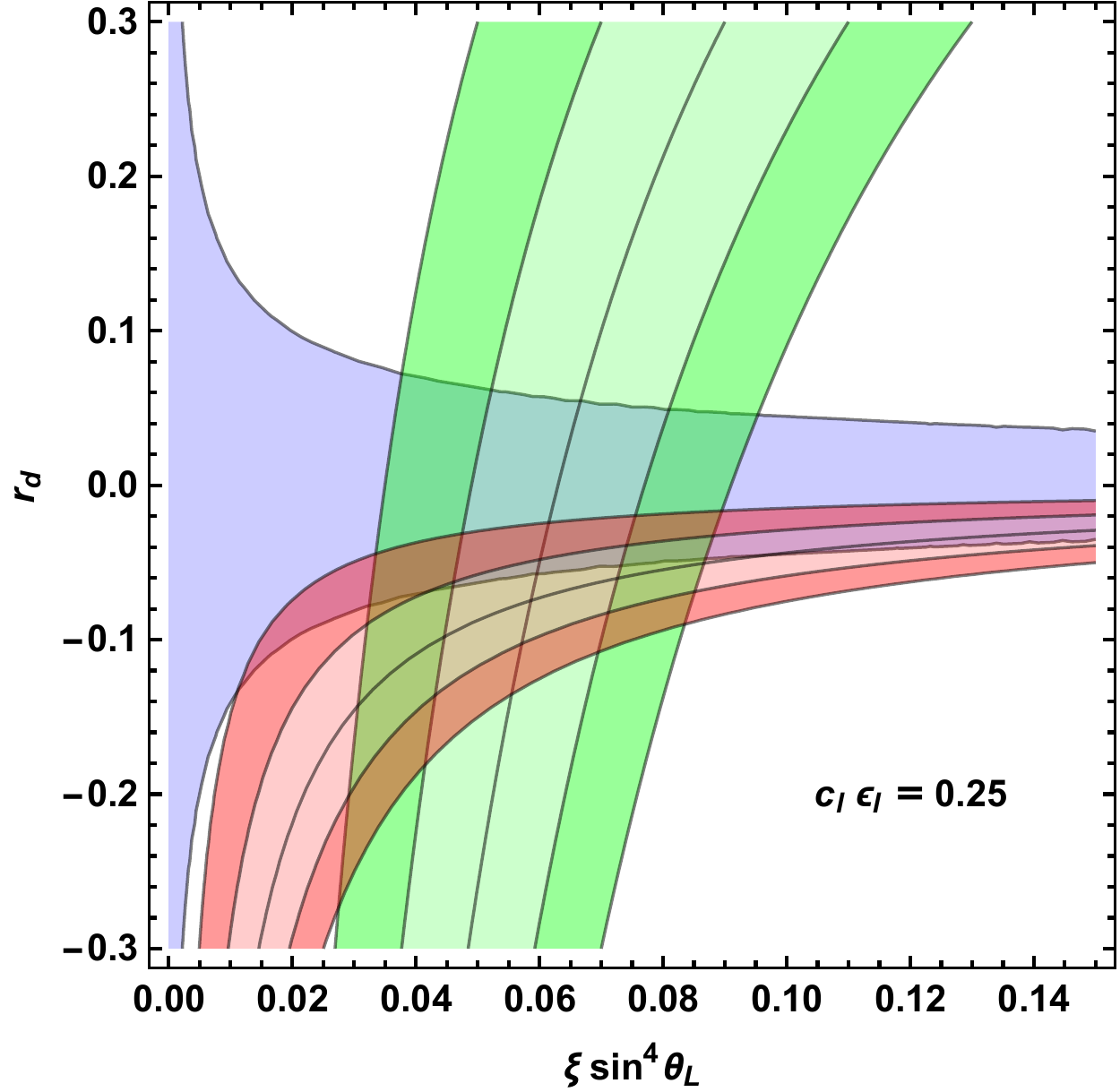}}
\caption{LEFT: Allowed parameter space for the neutral current anomaly, $R_K$, with $\xi=
\{0.075, 0.100, 0.125\}$. 
Dotted (full) contours delimit the 1$\sigma$ (2$\sigma$) regions. 
RIGHT: Allowed parameter space from $\Delta B_s=2$ (blue), $R_{D^{(*)}}$ (green) and
$R_K$ (red) for $s_{Lu3}=s_{L\nu 3}\equiv \sin{\theta_L}$ and $c_l \epsilon_l = 0.25$. Lighter (darker) regions are allowed at $1 \sigma $ ( $2 \sigma$). In both plots vector mixing angles have been neglected.}
  \label{fig3}
\end{figure}

Tree level $\Delta F = 2$ transitions are mediated by composite gluons $G^H$, and 
composite electroweak vectors $W^{H3}$, $\tilde{V}$, and $X$. In particular, for $\Delta B_s =2$ one has
\begin{align}
\label{Leff_bs}
\mathcal{L}_{eff}^{\Delta B_s = 2} &= 
r^2_d (V_{ts}V_{tb})^2
\Big(-\frac{g_2^2}{M_W^2}\Big)[\xi s^4_{Lu3}] \\
&\times \Big(\frac{1}{2}+\frac{1}{3}f_{G^H}+
\frac{1}{4}f_{W^*}+\frac{1}{36}\frac{3}{2}f_X\Big)
(\bar{s}_L\gamma_\mu b_L)^2 , \nn
\end{align}
where the $f$ functions are the same as in~\eqref{eq:f} with $\theta_{L\nu3}\rightarrow \theta_{Lu3}$, and $f_{G^H}=f_{W^*}$ with $\theta_2 \rightarrow \theta_3$. 
Neglecting the vector mixing angles in the $f$ functions, as previously done, $\Delta B_s=2$ data require
\begin{equation}
 r^2_d s^4_{Lu3} \Big(\frac{\xi}{0.1}\Big)\lesssim 2 \cdot 10^{-3} .
\end{equation}
The plots in fig.~\ref{fig3} show the parameter space needed to explain $R_{D^{(*)}}$ and $R_K$ as well as the parameter space consistent with measurements of $\Delta B_s = 2$ processes.\footnote{
At tree level, contributions to $b\rightarrow s \nu \bar{\nu}$ and $s\rightarrow d \nu \bar{\nu}$ are also present, mediated in the $s$-channel by the composite electroweak vectors $W^{H3}$, $\tilde{V}$, and $X$. 
They exhibit a CKM suppression proportional to $ r_d V_{tb}V^*_{ts}$ and $ r_d^2 V_{ts}V^*_{td}$  respectively.
The constraint from $\Delta B_s=2$ makes the corresponding experimental bounds irrelevant in the range of parameters considered in figure~\ref{fig3}.
}
The range of interest, mostly determined by $R_{D^{(*)}}$ and $\Delta B_s=2$, is 
\begin{equation}
0.03 \lesssim \xi s^2_{Lu3}s^2_{L\nu 3}\lesssim 0.09,\quad\quad\quad
-0.08 \lesssim r_d \lesssim -0.02 ,
\end{equation}
with the lower or upper bounds being reached simultaneously. The range of $r_d$ requires a tuning of the parameter $x_b= r_d/(1+r_d)$.

\section{Electroweak Precision Constraints}
\label{sec:ewpt}
Apart from flavor, composite Higgs models are constrained by electroweak precision data, which include oblique corrections, modified $Z$-couplings and modified right-handed $W$-couplings. 
Let us briefly comment on the new effects that are specific of the extension of $SU(3)$ to $SU(4)$ in the global symmetry group of the composite sector, related to the presence of the vector leptoquark. 

In the oblique corrections at one-loop no new contributions arise to the $S, T, U$ parameters from  exchanges of the  leptoquark, which is a singlet of $SU(2)_L$ (as is the case for all the neutral composite vectors). The only effect of the leptoquark is a contribution to the $Y$ parameter~\cite{Barbieri:2004qk}, which is UV-sensitive and will have to be cutoff at a scale $\Lambda$ by the composite dynamics. 
From the Lagrangian~\eqref{eq:LagV} one obtains~\cite{Barbieri:2015yvd}
\begin{equation}
Y= \frac{g_1^2}{64 \pi^2}\Big(\frac{4}{9}\Big)\frac{M_W^2}{M_{LQ}^2}\frac{\Lambda^2}{M_{LQ}^2}\approx
10^{-3}\Big(\frac{\xi}{0.1}\Big)\Big(\frac{1}{g_\rho^4}\Big) \Big(\frac{\Lambda}{4\pi f}\Big)^2,
\end{equation}
which does not pose any significant constraint on $g_\rho$ or $\xi$, at least compared with the usual constraints from $S$ and $T$. 
In this respect, in principle, a more important but also uncertain effect could come from the exchange of elementary/composite fermions, especially in view of the large mixing angles $s_{Lu3}$ and $s_{L\nu 3}$  needed to explain $R_{D^{(*)}}$ and $R_K$.  (See ref.s~\cite{Grojean:2013qca, Ghosh:2015wiz} for calculations in the global $SU(3)$ case).

Turning now our attention to non-oblique corrections, in particular to modifications of the $Z$-couplings, well known symmetry arguments~\cite{Agashe:2006at} imply a sufficient suppression of tree level corrections to  $Zbb$ and $Z\tau\tau$, $\delta g_{Lb}$ and $\delta g_{L\tau}$ respectively,  in the bidoublet model for  $z_3, z_{3e} \gtrsim 10$ or their exact absence in the triplet model. 
New effects, however, appear at one-loop where leptoquarks give rise to quadratically divergent effects in the 3-point function between $B_\mu$ and third generation fermions. Along the same lines of~\cite{Barbieri:2015yvd}, by considering one-loop diagrams with leptoquarks and SM fermions one obtains in both the bidoublet and triplet models
\begin{align}
 \delta g_{Lb} &= -\frac{1}{3}\frac{g_\rho^2}{64 \pi^2}\frac{s_W^2 M_{Z}^2}{M_{LQ}^2}\frac{\Lambda^2}{M_{LQ}^2} s_{Lu3}^2 s_{L\nu 3}^2
 \approx - 2 \cdot 10^{-3} \Big(\frac{\xi}{0.1}\Big)\Big(\frac{s_{Lu3}^2 s_{L\nu 3}^2}{g_\rho^2}\Big)\Big(\frac{\Lambda}{4\pi f}\Big)^2, \\
\delta g_{L\tau} &= \frac{g_\rho^2}{64 \pi^2}\frac{s_W^2 M_{Z}^2}{M_{LQ}^2}\frac{\Lambda^2}{M_{LQ}^2} s_{Lu3}^2 s_{L\nu 3}^2
  \approx  6 \cdot 10^{-3} \Big(\frac{\xi}{0.1}\Big)\Big(\frac{s_{Lu3}^2 s_{L\nu 3}^2}{g_\rho^2}\Big)\Big(\frac{\Lambda}{4\pi f}\Big)^2 . \nn
\end{align}
The strongest bound comes from $\delta g_{L\tau}$, enhanced by a color factor of 3 with respect to $\delta g_{Lb}$, which requires $g_\rho > 2\div 3$ for $\Lambda$ close to maximal.

\section{LHC phenomenology}
\label{sec:dist}
In the following we outline possible  features of the model relevant to LHC searches and distinctive with respect to the usual $SU(3) \times SO(5) \times U(1)$  minimal CHM setup.
In general there are three such features:
\begin{itemize}
\item There are a number of $Z^{\prime}$-like composite vector bosons, $\tilde{V}_{\mu},\, W_{\mu}^{H3},\, X_{\mu},\, B^H_{\mu}$, all of which, except $B^H_{\mu}$, have a large coupling to the left-handed components of the third  generation fermions. At LHC, by their  exchange in the $s$-channel, this results in a  significant effect in $b \bar{b}  \to \tau^+ \tau^-$.
\item There is a vector singlet composite leptoquark, $V_\mu$, partly responsible for the anomalies in $B$-decays, with branching ratios likely close to $50\%$ for $b\tau$ and $t\nu_\tau$. $V_\mu$ can be directly searched in QCD pair production. Its exchange in the $t$-channel also contributes to $b \bar{b}  \to \tau^+ \tau^-$.
\item There are exotic composite leptons with a mass within a few $\%$ degenerate with the exotic composite quarks that are normally discussed in the context of standard CHMs.
\end{itemize}

Before discussing in some details the first item, let us briefly comment  on the second item. To the best of our knowledge so far there has been only one search for the pair-production of spin-one leptoquarks decaying to third generation fermions at the LHC.
The CMS collaboration was able to set a bound of $M_V = M_{\rho} > 762$~GeV using 7~TeV data, assuming $\text{Br}(V_{-4/3} \to b \tau^-) = 100\%$ and that the leptoquarks are Yang-Mills-like ($k_s = 1$)~\cite{Chatrchyan:2012sv}.
This bound normally applies to the case $\text{Br}(V_{+2/3} \to b \tau^+) = 100\%$ as well, since the final state is $b \bar{b} \tau^- \tau^+$ in both cases.

Potentially stronger bounds on vector leptoquark masses might be obtained by reinterpreting scalar leptoquark pair-production searches that used 8 or 13~TeV data.
Table~\ref{tab:LQ} summarizes the experimental results relevant in this context. In the rightmost columns of tab.~\ref{tab:LQ}, checkmarks indicate which decays are in principle possible for a spin-one leptoquark that transforms as $L_Y$ under $SU(2)_L \times U(1)_Y$. $L_Y= 1_{2/3}$ is the case relevant in this context. By a simple reinterpretation of the results presented in ref.s~\cite{Khachatryan:2014ura, Khachatryan:2015bsa, Aad:2015caa, CMS:2016xxl},
we find that the bounds on spin-one leptoquark masses are currently limited by the range of masses considered in the experimental searches.
Due to this limitation, we urge the experimental collaborations to extend their search regions to include masses greater than 1~TeV.
%
\begin{table}
\centering
 \begin{tabular}{| c | c | c | c || c | c | c | c | c |}
 \hline 
Ref. & $\sqrt{s}$ (TeV) & Decay Mode & $M_V$ Range (GeV) & $1_{2/3}$ & $3_{2/3}$ & $2_{-5/6}$ & $1_{5/3}$ & $2_{1/6}$ \\ \hline \hline
\cite{Khachatryan:2014ura} & 8 & $V \to b \tau$ & $[200,\, 870]$ & \checkmark & \checkmark & \checkmark & & \\ \hline
\cite{Khachatryan:2015bsa} & 8 & $V \to t \tau$ & $[200,\, 800]$ & & \checkmark & \checkmark & \checkmark & \checkmark \\ \hline
\multirow{2}{*}{\cite{Aad:2015caa}} & \multirow{2}{*}{8} & $V \to b \nu_{\tau}$ & \multirow{2}{*}{$[200,\, 800]$} & & \checkmark & \checkmark & & \\
 & & $V \to t \nu_{\tau}$ &  & \checkmark & \checkmark & & & \checkmark \\ \hline
\cite{CMS:2016xxl} & 13 & $V \to b \tau$ & $[600,\, 1000]$ & \checkmark & \checkmark & \checkmark & & \\ \hline
 \end{tabular}
  \caption{Summary of experimental results on searches  for pair-production of scalar leptoquarks. 
In th columns on the right a checkmark indicates that the corresponding decay is in principle possible for a spin-one leptoquark that transforms as $L_Y$ under $SU(2)_L \times U(1)_Y$.}
  \label{tab:LQ}
\end{table}

\subsection{Resonances in $\tau^- \tau^+$}
\label{sec:taures}

Ref.~\cite{Buttazzo:2016kid} was the first to point out that $b \bar{b}  \to \tau^+ \tau^-$ can be a signal of models that attempt to explain $R_{D^{(*)}}$.
Subsequently ref.~\cite{Faroughy:2016osc} throughly investigated  bounds on explanations of $R_{D^{(*)}}$ coming from new physics searches involving pairs of tau leptons.
We will compare our results to those of~\cite{Faroughy:2016osc} at the end of this subsection.

\begin{figure}
  \centering
\includegraphics[width=0.6\textwidth]{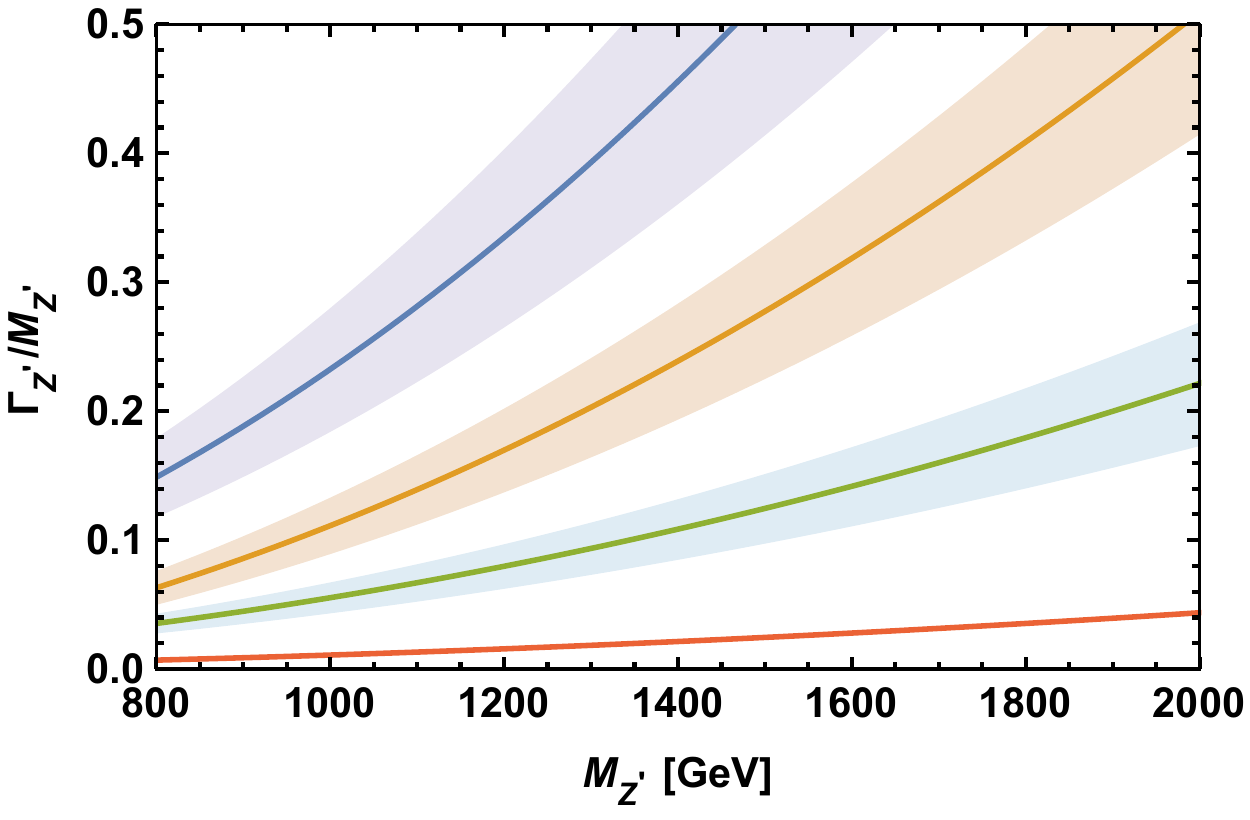}
 \caption{$\Gamma_{Z^{\prime}} / M_{Z^{\prime}}$ as a function of $M_{Z^{\prime}}$. 
 The blue, orange, green, and red curves correspond to $\tilde{V}_{\mu}$, $W_{\mu}^{H3}$, $X_{\mu}$, and $B^H_{\mu}$, respectively. 
 The solid lines reproduce the central value of $R_{D^{(*)}}$ assuming $s_{Lu3} = s_{L\nu 3}$ and $\xi = 0.1$, and the shaded bands reproduce $R_{D^{(*)}}$ at the $1\sigma$ level.}
  \label{fig:GammaMZprime}
\end{figure}

Let us first consider $Z^{\prime}$s alone.
The total width of the $Z^{\prime}$ generally includes decays to both pairs of third generation SM fermions as well as $W_L^+ W_L^-$ and $Z_L h$.
Decays to composite fermions, if allowed at all, are  phase space suppressed, and we neglect them in what follows.
Fig.~\ref{fig:GammaMZprime} shows $\Gamma_{Z^{\prime}} / M_{Z^{\prime}}$ as a function of $M_{Z^{\prime}}$. 
All of the relevant formulas can be found in appendix~\ref{sec:bbtautau}.
The blue, orange, green, and red curves correspond to $\tilde{V}_{\mu}$, $W_{\mu}^{H3}$, $X_{\mu}$, and $B^H_{\mu}$, respectively. 
The solid lines reproduce the central value of $R_{D^{(*)}}$ assuming $s_{Lu3} = s_{L\nu 3}$ and $\xi = 0.1$, and the shaded bands reproduce $R_{D^{(*)}}$ at the $1\sigma$ level.
We have not included a band for $B^H$ since its coupling to SM fermions in proportional to $t_R^2$, which leads to a feeble coupling.
%

The result of ATLAS at 8~TeV is the most constraining published experimental result on searches for new physics in $\tau^+ \tau^-$~\cite{Aad:2015osa}.
CMS has released preliminary results at 13~TeV that are na\"{i}vely more constraining for $M_{Z^{\prime}} \gsim 900$~GeV~\cite{CMS:2016zxk}.
However not enough information about cuts and efficiencies is provided to reinterpret this search in terms of the $Z^{\prime}$s of our model, which are not SM-like.
We used the publicly available plot digitizer WebPlotDigitizer v3.10~\cite{WPD310} in performing this analysis.

\begin{figure}
  \centering
\includegraphics[width=0.6\textwidth]{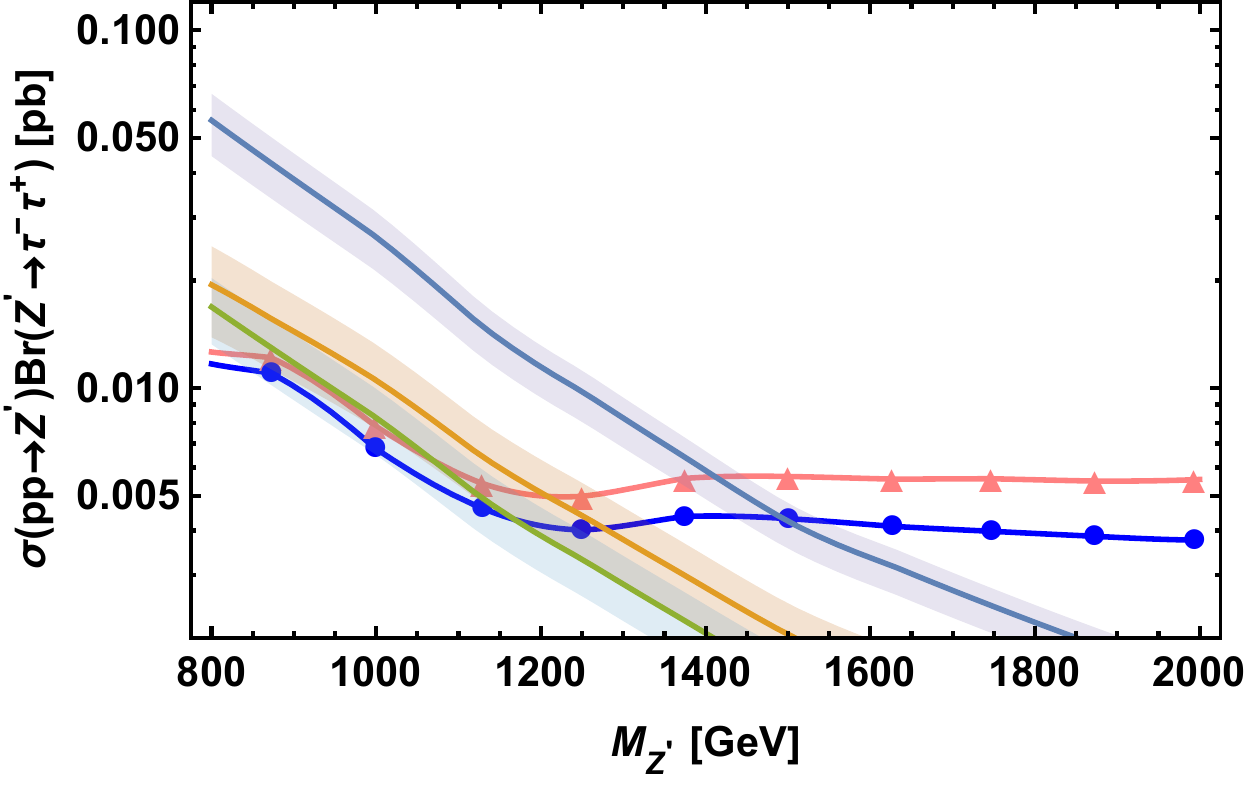}
 \caption{$\sigma(p p \to Z^{\prime} \to \tau^- \tau^+)$ versus $M_{Z^{\prime}}$ assuming the cross section is dominated by a single $Z^{\prime}$ and its interference with the SM. The blue, orange, and green curves correspond to $\tilde{V}_{\mu}$, $W_{\mu}^{H3}$, and $X_{\mu}$, respectively. }
  \label{fig:xsecZprime}
\end{figure}

Fig.~\ref{fig:xsecZprime} shows $\sigma(p p \to Z^{\prime} \to \tau^- \tau^+)$ versus $M_{Z^{\prime}}$ assuming the cross section is dominated by a single $Z^{\prime}$ and its interference with the SM.
The purely SM contribution is not included in fig.~\ref{fig:xsecZprime}.
The relevant formulas can again be found in appendix~\ref{sec:bbtautau}.
Throughout this work we use the NNPDF collaboration's \texttt{NNPDF23\_lo\_as\_0119} parton distribution function (PDF) grid~\cite{Ball:2012cx, Hartland:2012ia}.
To approximately match the cuts employed in the ATLAS search to those defined in eq.~\eqref{eq:btauhad} we take $Y_{\text{cut}} = 2.47$, and $p_{T\text{cut}} = m_T^{\text{tot}} / 2$ where $m_T^{\text{tot}}$ is defined in~\cite{Aad:2015osa} for a given $M_{Z^{\prime}}$.
The blue, orange, and green curves correspond to $\tilde{V}_{\mu}$, $W_{\mu}^{H3}$, and $X_{\mu}$, respectively. 
We have not shown the $B_{\mu}^H$ limit, as the corresponding cross section is proportional to $s_R^4$, which essentially yields no bound. 
Just as in fig.~\ref{fig:GammaMZprime}, the solid lines reproduce the central value of $R_{D^{(*)}}$ assuming $s_{Lu3} = s_{L\nu 3}$ and $\xi = 0.1$, and the shaded bands reproduce $R_{D^{(*)}}$ at the $1\sigma$ level.
The pink triangles and blue circles corresponds to the ATLAS upper limit on the cross section under two different assumptions about the nature of the $Z^{\prime}$, neither of which exactly corresponds to the $Z^{\prime}$s of this model. The pink triangle case assume SM-like couplings of the $Z^{\prime}$ to SM fermions, and an artificial width of 20\%. Whereas the blue circle case assumes only left-handed couplings to SM fermions, and the natural width that results from those couplings.
Agreement with the ATLAS search  requires at $1 \sigma$: $M_{\tilde{V}} \gsim 1.4$~TeV, $M_{W^H} \gsim 1.2$~TeV, and $M_X \gsim 1.1$~TeV.
For $X$ and $W^{H3}$ these limits are robust, as their widths are less than and approximately equal to 20\% of their mass at the implied limit, respectively. This is not the case for $\tilde{V}$, indicating that this limit is more uncertain.
%

Including only a single $Z^{\prime}$ at a time may not be a good approximation to the full BSM contribution to $\tau^- \tau^+$ production.
Using the relevant formulas in appendix~\ref{sec:bbtautau}, fig.~\ref{fig:xsectautau} shows a few scenarios where the complete contribution to tau pair production is included.
Specifically we plot $\sigma(p p \to \tau^- \tau^+)$ in invariant mass bins of 125~GeV versus $M_{\tau\tau}$.
The colored lines include all four $Z^{\prime}$s, the leptoquarks, the SM, and their interference.
The blue, orange, green, and red lines correspond to $\{M_V,\, M_{W^H}\} = \{1.0,\,1.5\},\, \{1.5,\,1.0\},\, \{1.0,\,1.0\},\, \{1.5,\,1.5\}$~TeV, respectively, with $M_V = M_X$ and $M_{W^H} = M_{\tilde{V}}$ in each case.
The black is the leading order pure SM contribution, which is not included in fig.~\ref{fig:xsecZprime}.
Once again the solid lines reproduce the central value of $R_{D^{(*)}}$ assuming $s_{Lu3} = s_{L\nu 3}$ and $\xi = 0.1$, and the shaded bands reproduce $R_{D^{(*)}}$ at the $1\sigma$ level. We have not attempted a detailed comparison of this cross section with the experimental data, which might exclude relevant regions of the $\{M_V,\, M_{W^H}\}$ plane but, we think, would still leave allowed points saturating the bounds obtained by the previous considerations.
%
\begin{figure}
  \centering
\includegraphics[width=0.6\textwidth]{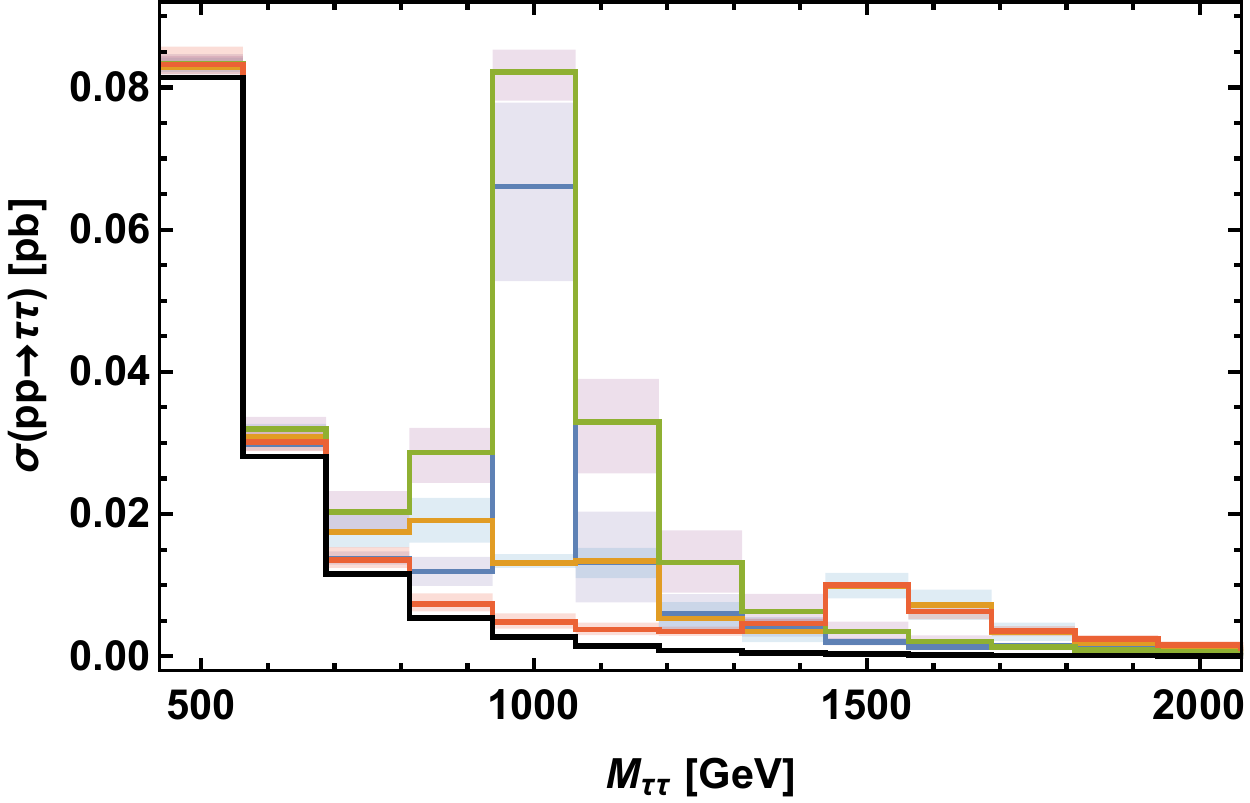}
 \caption{$\sigma(p p \to \tau^- \tau^+)$ in invariant mass bins of 125~GeV versus $M_{\tau\tau}$. 
The colored lines include all four $Z^{\prime}$s, the leptoquarks, the SM, and their interference, while the black is the SM alone. 
The blue, orange, green, and red lines correspond to $\{M_V,\, M_{W^H}\} = \{1.0,\,1.5\},\, \{1.5,\,1.0\},\, \{1.0,\,1.0\},\, \{1.5,\,1.5\}$~TeV, respectively, with $M_V = M_X$ and $M_{W^H} = M_{\tilde{V}}$ in each case.}
  \label{fig:xsectautau}
\end{figure}

By looking at individual $Z^{\prime}$ contributions, ref.~\cite{Faroughy:2016osc} found that $\tau^- \tau^+$ searches rule out some region of the space characterized by $(M_{Z^{\prime}}, \Gamma_{Z^{\prime}}, | g_{b} g_{\tau}| v^2/M_{Z^{\prime}}^2)$, where $\Gamma_{Z^{\prime}}$ is the total width and $g_b, g_\tau$ are the couplings of the $Z^{\prime}$ to the $b$ and the $\tau$.
For the $Z^{\prime}$s, a direct comparison is possible for our $\tilde{V}, W^H$ and $X$ bosons, using the total widths given in appendix~\ref{sec:bbtautau} and the approximate relations
\begin{equation}
\label{eq:Zprimecomp}
\frac{\left| g_{\tilde{V} b} g_{\tilde{V} \tau}\right| v^2}{M_{\tilde{V}}^2} = 2 \frac{\left| g_{W^H b} g_{W^H \tau}\right| v^2}{M_{W^H}^2} = 4 \frac{\left| g_{X b} g_{X \tau}\right| v^2}{M_{X}^2} = \xi s_{Lu3}^2 s_{L\nu3}^2 .
\end{equation}
We find good agreement between our results and the ones derivable from the figure~4 of~\cite{Faroughy:2016osc}.\footnote{Note that we use $v \approx 175$~GeV, whereas in \cite{Faroughy:2016osc} one uses $v \approx 250$~GeV.}

We did not directly investigate the bounds on the leptoquark of this work coming from its  sole contribution to $\tau^- \tau^+$ searches.
However it is straightforward to translate the results of~\cite{Faroughy:2016osc} into the parameters of our model.
The composite leptoquarks have the same couplings structure as the leptoquarks in the so-called minimal model.
The difference in terms on bounds is that in the composite model, $R_{D^{(*)}}$ receives approximately equal contributions from the leptoquark and the $W^{H3}$ boson.
Thus in bounding the leptoquark parameter one should rescale the parameter $g_U$ of~\cite{Faroughy:2016osc} by a factor of $1 / \sqrt{2}$.
In doing so, the bounds on vector leptoquark from $\tau^- \tau^+$ searches in the upper panel of figure~6 of~\cite{Faroughy:2016osc} are relaxed.

\section{Conclusions}
\label{sec:conc}

Measurements in flavor physics in the years to come can compete with direct searches at the LHC in the attempt to discover deviations from the SM. In our view this is especially the case if a weakly broken $U(2)^n$  symmetry plays some role in determining the structure of flavor. Given this premise it is natural to give consideration to a number of anomalies emerging in the decays of $B$ mesons. On one side there is the fact that the statistically most significant among these anomalies, $b\rightarrow c \tau \bar{\nu}$,  involves three third generation particles. This matches with a $U(2)^n$, which brings in a basic distinction between the third generation and the first two lighter families. On the other side there is the  relatively large size of the putative deviation from a SM tree level amplitude, making it difficult to conceive a purely perturbative interpretation. 

Building on these considerations and based on ref.~\cite{Barbieri:2015yvd}, in this work we have attempted to construct a partially UV complete explanation of the putative anomalies by extending the minimal CHM  from an $SU(3)\times SO(5)\times U(1)$ to an $SU(4)\times SO(5)\times U(1)$ global symmetry group. With a suitable choice of the representation of the composite fermions under the unbroken global symmetry group we have shown that such construction can be performed without manifest contradiction with current experiments. To explain the anomalies requires large mixings of the left-handed third-generation quarks and leptons, $| s_{Lu3}s_{L\nu 3}|\approx 0.7\div 0.8$
for $\xi=0.1$ and, to be consistent with  $b \bar{b}  \to \tau^+ \tau^-$ searches at LHC, relatively large couplings of the composite vectors, $g_{\rho}, g_{\rho R} \gtrsim 3\div 4$, always for $\xi=0.1$.

In case the anomalies will persist and perhaps be reinforced in experiments to come, several more detailed investigations can be performed of the model described here to prove its full compatibility with the various constraints, present and future. They include three main chapters: i) electroweak corrections, both of oblique and non-oblique nature, extending and completing sec.~\ref{sec:ewpt}; ii) flavor physics, as partly already discussed in \cite{Barbieri:2015yvd}, both in the quark and in the lepton sector; iii) LHC searches, as outlined in sec.~\ref{sec:dist}.

\begin{acknowledgments}
We thank Oleksii Matsedonskyi for useful discussions, David Straub for useful discussions and comments on the manuscript, and the Institute of Theoretical Studies at ETH Z\"{u}rich for its hospitality while part of this work was completed. 
The work of RB was supported in part by the Swiss National Science Foundation under contract 200021-159720. 
The work of CM was supported in part by the Italian Ministry of Education, University and Research's Fund for Investment in Basic Research under grant RBFR12H1MW, and by the United States Department of Energy under grant contract de-sc0012704. 
\end{acknowledgments}

\appendix

\section{Triplet Scenario}
\label{sec:trip}
The discussion for the $SU(4)$ extension of the triplet scenario proceeds along the same lines as the bidoublet scenario: the fermionic particle content is
\begin{equation}
\psi = 
\begin{pmatrix}
Q^{\beta} \\
L
\end{pmatrix}, \quad
\chi = 
\begin{pmatrix}
\chi^{\beta}_{\tilde{Q}} \\
\chi_{\tilde{L}}
\end{pmatrix}, \quad
\chi^{\prime} = 
\begin{pmatrix}
\chi^{\prime \beta}_{\tilde{Q}} \\
\chi^{\prime}_{\tilde{L}}
\end{pmatrix} ,
\end{equation}
where $\beta = 1,2,3$ is a fundamental color index. 
Under $SU(4)\times SU(2)_L \times SU(2)_R \times U(1)_X$ they transform like $\psi=(4,2,2)_{ 1/2}$, $\chi=(4,1,3)_{ 1/2}$ and $\chi^{\prime}=(4,3,1)_{ 1/2}$. 
The components of $\psi$, $\chi$ and $\chi^{\prime}$ are
\begin{equation}
Q =
\begin{pmatrix}
Q^{(U)} & X^{(U)}
\end{pmatrix}
=
\begin{pmatrix}
U & X^{5/3} \\
D & X^{2/3}
\end{pmatrix}, \quad\quad
L =
\begin{pmatrix}
L^{(N)} & X^{(N)}
\end{pmatrix}
=
\begin{pmatrix}
N & X^{+1} \\
E & X^{0}
\end{pmatrix}, 
\end{equation}
\begin{equation}
\chi_{\tilde{Q}} =
\begin{pmatrix}
\tilde{X}^{5/3}\\
\tilde{U}\\
\tilde{D}
\end{pmatrix}, \quad\quad
\chi_{\tilde{L}} =
\begin{pmatrix}
\tilde{X}^{+1}\\
\tilde{N}\\
\tilde{E}
\end{pmatrix}, \quad\quad
\chi^{\prime}_{\tilde{Q}} =
\begin{pmatrix}
\tilde{X}^{\prime 5/3}\\
\tilde{U}^{\prime}\\
\tilde{D}^{\prime}
\end{pmatrix}, \quad\quad
\chi^{\prime}_{\tilde{L}} =
\begin{pmatrix}
\tilde{X}^{\prime +1}\\
\tilde{N}^{\prime}\\
\tilde{E}^{\prime}
\end{pmatrix} .
\end{equation}
A notable difference is that there is only one composite doublet with the same quantum numbers of $q_L$, consequently $z_3=z_{12}=1$. 
This is not a problem for the $Zb_L \bar{b}_L$ coupling deviation because with this choice of representations for composite quarks the tree level deviation is zero as can be understood by the symmetry considerations of \cite{Agashe:2006at}.
Another important difference is that composite states with the same quantum numbers of $u_R$ and $d_R$ are inside the $(1,3)$ multiplet of $SO(4)\cong SU(2)_L\times SU(2)_R$ and do not live in different multiplets as in the bidoublet case. 
Therefore, right-compositeness cannot be implemented and we will focus on left compositeness
\begin{equation}
\mathcal{G}_{LC}=U(2)_{q+l+\psi+\chi+\chi'}\times U(2)_u \times
U(2)_d\times U(2)_e .
\end{equation}

The Yukawa and mass terms for the fermionic resonances in the strong sector are given by
\begin{equation}
\begin{split}
\mathcal{L}^{\text{triplet}}_{s}=
m_{\psi}\text{Tr}[\bar{\psi} \psi]+
m_{\chi}\text{Tr}[\bar{\chi} \chi]+m_{\chi^{\prime}}\text{Tr}[\bar{\chi}^{\prime}\chi^{\prime}]+\\
\Big(
Y_+^{ii}\, \text{Tr}[\bar{\psi}^i_L \mathcal{H}\chi^i_R] + 
Y_-^{ii}\, \text{Tr}[\mathcal{H}\bar{\psi}_L^i \chi'^i_R]+ \text{h.c.} \Big) ,
\end{split}
\end{equation}
where $Y_{\pm}$, $m_{\psi}$ and $m_{\chi^{(')}}$ are $U(2)$ preserving flavor diagonal matrices. 
Quark and lepton mixing Lagrangians are given by
\begin{align}
\label{trmixing2L} 
 \mathcal{L}_\text{q mix,LC}^{\text{triplet}} &=
m_{\psi_{3}}\lambda_{Lq3}\bar{q}_{L3}Q^U_{3R} +
m_{\psi_{2}}\lambda_{Lq2} \mathbf{\bar{q}_L} \mathbf{Q^U_R} +
m_{\chi_{3}}\lambda_{Ru3}\tilde{T}_L t_R
\notag\\
&+m_{\chi_{2}}\,d_u\, (\bar{\tilde{\mathbf{U}}}_\mathbf{L}\V)t_R  +
m_{\chi_{2}}\, \bar{\tilde{\mathbf{U}}}_\mathbf{L} \Delta_u \bar{\mathbf{u}}_{\mathbf{R}} +
\text{h.c.}
+ (u,\tilde{U},t,\tilde{T}\rightarrow d,\tilde{D},b,\tilde{B}),
\end{align}
\begin{align}
\label{trmixinge}
 \mathcal{L}_\text{e mix,LC}^{\text{triplet}} &=
m_{\psi_{3}}\lambda_{Le3}\bar{l}_{3L}L^N_{3R} +
m_{\psi_{2}}\lambda_{Le2} \bar{\mathbf{l}}_{\mathbf{L}}\mathbf{L^N_R} +
m_{\chi_{3}}\lambda_{Re3}\bar{\tilde{E}}_{3L}\tau_R
\notag\\
&+m_{\chi_{2}}\,d_e\, (\bar{\tilde{\mathbf{E}}}_\mathbf{L}\V)\tau_R +
m_{\chi_{2}}\bar{\tilde{\mathbf{E}}}_\mathbf{L} \Delta_e \bar{\mathbf{e}}_{\mathbf{R}} +
\text{h.c.} .
\end{align}
The SM Yukawa matrices can be written as in eq.s~\eqref{SMYukLC} and~\eqref{SMYukLCe} with
\begin{align}
y_t &= Y_{+3}s_{Lq3}s_{Ru3},
\\
a_u &= Y_{+2}s_{Lq2},& x_b &=\Big(\frac{d_d Y_{-2}Y_{+3}}{d_u  Y_{-3}Y_{+2}}\Big)
\frac{t_{Ru3}}{t_{Rd3}}
, & &  \\
y_\tau &= Y_{-3}s_{Le3}s_{Re3},
\\
a_e &= Y_{-2}s_{Le2},& x_\tau &=\Big(\frac{d_e Y_{-2}Y_{+3}}{d_u  Y_{-3}Y_{+2}}\Big)
\frac{s_{Le2}s_{Lq3}t_{Ru3}}{s_{Lq2}s_{Le3}t_{Re3}} ,
\end{align}
with $y_b$ and $a_d$ given by the same expressions of $y_t$ and $a_u$  provided that right up-mixing angles and $Y_+$ are replaced by right down-mixing angles and $Y_-$.


\section{Interactions of Composite Vectors with Elementary Fermions}
\label{sec:vecferm}
This appendix gives the interaction Lagrangians of composite vectors with elementary quarks and leptons that are relevant for the processes discussed in section~\ref{tran}.
Each piece of the Lagrangian below contains one composite vector. 
Explicitly, the individual terms are:
\begin{equation}
 \mathcal{L}_{HG} = g_\rho c_3 s^2_{Lu3} A^{(3)}_{Lu3} G_{\mu}^{H a}\Big
 [U^{d*}_{3i}U^d_{3j}\bar{d}_{Li}\gamma^\mu \frac{\lambda^a}{2} d_{Lj}+
 U^{u*}_{3i}U^u_{3j}\bar{u}_{Li}\gamma^\mu \frac{\lambda^a}{2} u_{Lj}\Big] ,
\end{equation}
\begin{equation}
 \mathcal{L}_{LQ} = \frac{g_\rho}{\sqrt{2}} s_{Lu3}s_{L\nu3} V_{\mu}\Big
 [\bar{d}_{L}\gamma^\mu F^D e_{L}+
 \bar{u}_{L}\gamma^\mu F^U \nu_{L}\Big]+\text{h.c.} ,
\end{equation}
\begin{equation}\begin{split}
 \mathcal{L}_{X} = \sqrt{\frac{3}{2}}g_{\rho} c_{15} X_{\mu}\Big
 [\frac {A^{(15)}_{Lu3}s_{Lu3}^2}{6}\Big(  U^{d*}_{3i}U^d_{3j}\bar{d}_{Li}\gamma^\mu  d_{Lj}+
 U^{u*}_{3i}U^u_{3j}\bar{u}_{Li}\gamma^\mu u_{Lj}\Big)-\\
 \frac{A^{(15)}_{L\nu3}s_{L\nu3}^2}{2}U_{3i}^{e*}U_{3j}^{e}\Big(
 \bar{e}_{Li}\gamma^\mu e_{Lj}+\bar{\nu}_{Li}\gamma^\mu \nu_{Lj}
 \Big)\Big] , \quad\quad
 \end{split}
\end{equation}
\begin{align}
 \mathcal{L}_{\tilde{V}} = -\frac{g_{\rho R}}{\sqrt{2}} \tilde{V}_{\mu}&\Big
 [s_{Lu3}^2\Big(  U^{d*}_{3i}U^d_{3j}\bar{d}_{Li}\gamma^\mu  d_{Lj}+
 U^{u*}_{3i}U^u_{3j}\bar{u}_{Li}\gamma^\mu u_{Lj}\Big) \\
 &+s_{L\nu3}^2U_{3i}^{e*}U_{3j}^{e}\Big(
 \bar{e}_{Li}\gamma^\mu e_{Lj}+\bar{\nu}_{Li}\gamma^\mu \nu_{Lj}
 \Big)\Big] , \nn
\end{align}
\begin{align}
 \mathcal{L}_{B^H} = -\frac{g_{\rho R}}{\sqrt{2}}c_R t_R^2 B^H_{\mu}&\Big
 [\frac{c_{Lu3}^2}{6}\Big(  U^{d*}_{3i}U^d_{3j}\bar{d}_{Li}\gamma^\mu  d_{Lj}+
 U^{u*}_{3i}U^u_{3j}\bar{u}_{Li}\gamma^\mu u_{Lj}\Big) \\
 &-\frac{c_{L\nu3}^2}{2}U_{3i}^{e*}U_{3j}^{e}\Big(
 \bar{e}_{Li}\gamma^\mu e_{Lj}+\bar{\nu}_{Li}\gamma^\mu \nu_{Lj}
 \Big)\Big] , \nn
\end{align}
\begin{equation}\begin{split}
 \mathcal{L}_{W^{H3}} = \frac{g_{\rho L}}{2}c_2 W^{H3}_{\mu}\Big
 [s_{Lu3}^2 A^{(2)}_{Lu3}
 \Big(  -U^{d*}_{3i}U^d_{3j}\bar{d}_{Li}\gamma^\mu  d_{Lj}+
U^{u*}_{3i}U^u_{3j}\bar{u}_{Li}\gamma^\mu u_{Lj}\Big)+\\
 s_{L\nu3}^2 A^{(2)}_{L\nu 3}U_{3i}^{e*}U_{3j}^{e}\Big(
 -\bar{e}_{Li}\gamma^\mu e_{Lj}+\bar{\nu}_{Li}\gamma^\mu \nu_{Lj}
 \Big)\Big] , \end{split}
\end{equation}
\begin{equation}
 \mathcal{L}_{W^{H\pm}} = \frac{g_{\rho L}}{\sqrt{2}}c_2 W^{H+}_{\mu}
 \Big[
 s_{Lu3}^2A^{(2)}_{Lu3}U^{u*}_{3i}U^d_{3j}\bar{u}_{Li}\gamma^\mu d_{Lj}+
 s_{L\nu 3}^2A^{(2)}_{L\nu3}U^{e*}_{i3}U^{e}_{3j}\bar{\nu}_{Li}\gamma^\mu e_{Lj}\big]
 +\text{h.c.} .
 \end{equation}
The coefficients $A^{(V)}_f$ are defined in eq.~\eqref{eq:Adef}, and the
matrix elements $U^{d,u}_{3i}$ can be written in terms of CKM matrix elements as
\begin{equation}
 U^d_{3i}=\left\{
                \begin{array}{ll}
                  r_d V_{ti} \quad i=1,2\\
                  V_{tb} \quad i=3
               \end{array}
             \right. ,
\quad\quad\quad
U^u_{3i}=\left\{
                \begin{array}{ll}
                  r_u V^*_{ib} \quad i=1,2\\
                  V_{tb} \quad i=3
               \end{array}
              \right. .
\end{equation}
The leptoquark Lagrangian, $\mathcal{L}_{LQ}$, makes use of the following definitions
\begin{equation}
F^U  = \begin{pmatrix}
-V_{ub}(s_l\epsilon_l)[1-r_d] &   -V_{ub}(c_l\epsilon_l)[1-r_d]& V_{ub}[1-r_d]\\     
-V_{cb}(s_l\epsilon_l)[1-r_d] &   -V_{cb}(c_l\epsilon_l)[1-r_d]& V_{cb}[1-r_d]\\ 
-V_{tb}(s_l\epsilon_l) &   -V_{tb}(c_l\epsilon_l)(b-1)& V_{tb}\\ 
\end{pmatrix} ,
\end{equation}
\begin{equation}
F^D = \begin{pmatrix}
-V_{td}(s_l\epsilon_l)r_d &   -V_{td}(c_l\epsilon_l)r_d& V_{td}r_d\\     
-V_{ts}(s_l\epsilon_l)r_d &   -V_{ts}(c_l\epsilon_l)r_d& V_{ts}r_d\\ 
-V_{tb}(s_l\epsilon_l) &  -V_{tb}(c_l\epsilon_l)& V_{tb}\\ 
\end{pmatrix} ,
\end{equation}
where $\theta_l$ is the angle ($s_l =\sin\theta_l$,  $c_l=\cos\theta_l$) in the unitary transformation which diagonalizes $\Delta_e$ on the left side and $\epsilon_l \equiv x_\tau |\bold{V}|$.

In the couplings of $V_\mu$ and $W_\mu^{H\pm}$ we are neglecting terms in the $1-2$ sector proportional to the square of the small $s_{L2}$ mixings.
The interactions of $W^{R \pm}$ are not shown since they always involve at least one exotic fermion.

\section{Formulas for Decay Rates and Cross Sections}
\label{sec:bbtautau}
In this appendix we give the formulas necessary to compute $\tau^- \tau^+$ pair production at the LHC, which in this model is initiated primarily by $b \bar{b}$.
The relevant couplings, which can be read off of the equations in appendix~\ref{sec:vecferm}, are given to a good approximation by
\begin{align}
g_{Xb} &= \frac{\sqrt{3}}{6} \frac{\sqrt{\xi} M_X}{v} s_{Lu3}^2, \quad g_{X\tau} = - \frac{\sqrt{3}}{2} \frac{\sqrt{\xi} M_X}{v} s_{L\nu3}^2,  \\
g_{W^{H3}b} &= - \frac{\sqrt{2}}{2} \frac{\sqrt{\xi} M_{W^H}}{v} s_{Lu3}^2, \quad g_{W^{H3}\tau} = - \frac{\sqrt{2}}{2} \frac{\sqrt{\xi} M_{W^H}}{v} s_{L\nu3}^2, \nn \\
g_{\tilde{V}b} &= - \frac{\sqrt{\xi} M_{\tilde{V}}}{v} s_{Lu3}^2, \quad g_{\tilde{V}\tau} =  \frac{\sqrt{\xi} M_{\tilde{V}}}{v} s_{L\nu3}^2 , \quad g_{Vb\tau} = \frac{\sqrt{\xi} M_V}{v} s_{Lu3} s_{L\nu3} . \nn 
\end{align}
Similarly the decay rates of the leptoquarks and the $Z^{\prime}$s are
\begin{align}
\Gamma_V &= \frac{M_V}{24 \pi} g_{\rho}^2 s_{Lu3}^2 s_{L\nu3}^2 , \quad
\Gamma_X = \frac{M_X}{96 \pi} g_{\rho}^2 \left(s_{Lu3}^4 + 3 s_{L\nu3}^4\right) , \quad
\Gamma_{B^H} = \frac{M_{B^H}}{192 \pi} g_{\rho R}^2 , \\
\Gamma_{W^{H3}} &= \frac{M_{W^H}}{96 \pi} g_{\rho L}^2 \left(1 + 6 s_{Lu3}^2 + 2 s_{L\nu3}^2\right) , \nn \quad
\Gamma_{\tilde{V}} = \frac{M_{\tilde{V}}}{192 \pi} g_{\rho R}^2 \left(1 + 24 s_{Lu3}^4 + 8 s_{L\nu3}^4\right) . \nn
\end{align}
Furthermore the BSM partonic level cross section for $b \bar{b} \to \tau^- \tau^+$ including interference with the SM, but ignoring the masses of the SM fermions is given by
\begin{align}
\label{eq:parton}
\frac{d\hat{\sigma}(b \bar{b} \to \tau^- \tau^+)}{d\hat{t}} &= \frac{1}{48 \pi} \frac{\hat{u}^2}{\hat{s}^2} \left[g_{Vb\tau}^4 \left|\Delta_V(\hat{t})\right|^2 \right. \\
&\left.+ \left(g_{Xb}^2 g_{X\tau}^2 \left|\Delta_X(\hat{s})\right|^2 + (X \to W^{H3}) + (X \to \tilde{V})\right)\right. \nn \\
&\left.+ 2 g_{Vb\tau}^2 \left(g_{Xb} g_{X\tau}\, \text{Re}\left(\Delta_V(\hat{t}) \Delta_X^*(\hat{s})\right) + (X \to W^{H3}) + (X \to \tilde{V})\right) \right. \nn \\
&\left.+ 2\left(g_{Xb} g_{X\tau} g_{W^{H3}b} g_{W^{H3}\tau}\, \text{Re}\left(\Delta_X(\hat{s}) \Delta_{W^{H3}}^*(\hat{s})\right) \right.\right. \nn \\
&\left.\left.+ (X \to \tilde{V}) + (W^{H3} \to \tilde{V})\right)\right] \nn \\
&+ \frac{\alpha}{6} \frac{\hat{u}^2}{\hat{s}^2}\left[g_{Vb\tau}^2 \left(Q_b Q_{\tau}\, \text{Re}\left(\Delta_V(\hat{t}) \Delta_{\gamma}(\hat{s})\right) + \frac{g_{Lb} g_{L\tau}}{s_W^2 c_W^2}\, \text{Re}\left(\Delta_V(\hat{t}) \Delta_Z^*(\hat{s})\right)\right) \right. , \nn \\
&\left.+ \left(g_{Xb} g_{X\tau}\left(Q_b Q_{\tau}\, \text{Re}\left(\Delta_X(\hat{s}) \Delta_{\gamma}(\hat{s})\right) + \frac{g_{Lb} g_{L\tau}}{s_W^2 c_W^2}\, \text{Re}\left(\Delta_X(\hat{s}) \Delta_Z^*(\hat{s})\right)\right) \right.\right. \nn \\
&\left.\left.+ (X \to W^{H3}) + (X \to \tilde{V})\right)\right] , \nn
\end{align}
with $Q_f$ and $g_{Lf}$ being the electric charge and left-handed coupling of the fermion $f$ to the $Z$, respectively.
In addition, we have defined $\Delta_X^{-1}(s) = s - M_X^2 + i \Gamma_X M_X$, etc..
To obtain hadronic level cross sections we first need the parton luminosity functions
\begin{align}
f\!\!f_{gg}\left(y, \mu_F^2\right) &= \int_y^1 \! \frac{dx}{x}\, f_{g / p}\left(x, \mu_F^2\right) f_{g / p}\left(\frac{y}{x}, \mu_F^2\right) , \\
f\!\!f_{q\bar{q}}\left(y, \mu_F^2\right) &= \int_y^1 \! \frac{dx}{x}\, \left(f_{q / p}\left(x, \mu_F^2\right) f_{\bar{q} / p}\left(\frac{y}{x}, \mu_F^2\right) + f_{\bar{q} / p}\left(x, \mu_F^2\right) f_{q / p}\left(\frac{y}{x}, \mu_F^2\right)\right), \nn
\end{align}
where $f_{i / p}$ is the PDF of species $i$, and $\mu_F$ is the factorization scale.
From this we can immediately write down the invariant mass distribution for $\tau^- \tau^+$ production
\begin{equation}
\frac{d\sigma\left(p p \to \tau^- \tau^+\right)}{dM_{\tau\tau}} = \frac{2 M_{\tau\tau}}{s} f\!\!f_{b\bar{b}}\left(\frac{M_{\tau\tau}^2}{s}, M_{\tau\tau}^2\right) \hat{\sigma}\left(b \bar{b} \to \tau^- \tau^+\right),
\end{equation}
with $s$ being the square of the collider center-of-mass energy, and $M_{\tau\tau}^2 = \hat{s}$.
The ATLAS search~\cite{Aad:2015osa} for new physics in $\tau^- \tau^+$ places cuts on the transverse momentum and rapidity of the tau leptons, which leads to the following form for the cross section\footnote{It is only necessary to impose the rapidity cut when $p_T^2 < s / (4 \cosh^2(Y_{\text{cut}}))$. For larger values of the transverse momentum, the rapidity is instead bounded by $|Y| < \text{arccosh}(\sqrt{s / (4 p_T^2)})$. Computationally it is somewhat faster to integrate over one region as in eq.~\eqref{eq:btauhad} as opposed to two regions. This is made possible because our implementation of the PDFs evaluate to zero whenever $x$ (or $y/x$) is greater than one.}
\begin{equation}
\label{eq:btauhad}
\sigma(p p \to \tau^- \tau^+) = \int_{p_{T\text{cut}}^2}^{s /4} \! dp_T^2 \int_{- Y_{\text{cut}}}^{Y_{\text{cut}}} \! dY \, \frac{2 \hat{s}}{s} f\!\!f_{b\bar{b}}\left(\frac{\hat{s}}{s}, p_T^2\right) \frac{d\hat{\sigma}(b \bar{b} \to \tau^- \tau^+)}{d\hat{t}} .
\end{equation}
Recall that for massless fermions $\hat{s} = 4 p_T^2 \cosh^2(Y)$ and $\hat{t} = - 2 p_T^2 \cosh(Y) e^{-Y}$.

\bibliographystyle{utphys}
\bibliography{v10h}

\end{document}